\documentclass[12 pt, amsfonts, amssymb,color]{article}

\usepackage{amsmath,amssymb,amsfonts,latexsym,float,graphics,epsfig,epstopdf}
\usepackage{subfig}
\usepackage{verbatim}

\begin{document}
%<<<<<<<<<<< enumeration of eqns section wise>>>>>>>>>>>>>>>>>>>

\renewcommand\theequation{\arabic{section}.\arabic{equation}}
\catcode`@=11 \@addtoreset{equation}{section}
%<<<<<<<<<<<<<<<<<<<<<<<<<<<<<<<<<>>>>>>>>>>>>>>>>>>>>>>>>>>>>>>>>>
\newtheorem{axiom}{Definition}[section]
\newtheorem{theorem}{Theorem}[section]
\newtheorem{axiom2}{Example}[section]
\newtheorem{lem}{Lemma}[section]
\newtheorem{prop}{Proposition}[section]
\newtheorem{cor}{Corollary}[section]
\newcommand{\be}{\begin{equation}}
\newcommand{\ee}{\end{equation}}
\newcommand{\lI}{\lambda_I}
\newcommand{\lR}{\lambda_R}
\newcommand{\nn}{\nonumber}

\newcommand{\equal}{\!\!\!&=&\!\!\!}
\newcommand{\rd}{\partial}
\newcommand{\g}{\hat {\cal G}}
\newcommand{\bo}{\bigodot}
\newcommand{\res}{\mathop{\mbox{\rm res}}}
\newcommand{\diag}{\mathop{\mbox{\rm diag}}}
\newcommand{\Tr}{\mathop{\mbox{\rm Tr}}}
\newcommand{\const}{\mbox{\rm const.}\;}
\newcommand{\cA}{{\cal A}}
\newcommand{\bA}{{\bf A}}
\newcommand{\Abar}{{\bar{A}}}
\newcommand{\cAbar}{{\bar{\cA}}}
\newcommand{\bAbar}{{\bar{\bA}}}
\newcommand{\cB}{{\cal B}}
\newcommand{\bB}{{\bf B}}
\newcommand{\Bbar}{{\bar{B}}}
\newcommand{\cBbar}{{\bar{\cB}}}
\newcommand{\bBbar}{{\bar{\bB}}}
\newcommand{\bC}{{\bf C}}
\newcommand{\cbar}{{\bar{c}}}
\newcommand{\Cbar}{{\bar{C}}}
\newcommand{\Hbar}{{\bar{H}}}
\newcommand{\cL}{{\cal L}}
\newcommand{\bL}{{\bf L}}
\newcommand{\Lbar}{{\bar{L}}}
\newcommand{\cLbar}{{\bar{\cL}}}
\newcommand{\bLbar}{{\bar{\bL}}}
\newcommand{\cM}{{\cal M}}
\newcommand{\bM}{{\bf M}}
\newcommand{\Mbar}{{\bar{M}}}
\newcommand{\cMbar}{{\bar{\cM}}}
\newcommand{\bMbar}{{\bar{\bM}}}
\newcommand{\cP}{{\cal P}}
\newcommand{\cQ}{{\cal Q}}
\newcommand{\bU}{{\bf U}}
\newcommand{\bR}{{\bf R}}
\newcommand{\cW}{{\cal W}}
\newcommand{\bW}{{\bf W}}
\newcommand{\bZ}{{\bf Z}}
\newcommand{\Wbar}{{\bar{W}}}
\newcommand{\Xbar}{{\bar{X}}}
\newcommand{\cWbar}{{\bar{\cW}}}
\newcommand{\bWbar}{{\bar{\bW}}}
\newcommand{\abar}{{\bar{a}}}
\newcommand{\nbar}{{\bar{n}}}
\newcommand{\pbar}{{\bar{p}}}
\newcommand{\tbar}{{\bar{t}}}
\newcommand{\ubar}{{\bar{u}}}
\newcommand{\utilde}{\tilde{u}}
\newcommand{\vbar}{{\bar{v}}}
\newcommand{\wbar}{{\bar{w}}}
\newcommand{\phibar}{{\bar{\phi}}}
\newcommand{\Psibar}{{\bar{\Psi}}}
\newcommand{\bLambda}{{\bf \Lambda}}
\newcommand{\bDelta}{{\bf \Delta}}
\newcommand{\p}{\partial}
\newcommand{\om}{{\Omega \cal G}}
\newcommand{\ID}{{\mathbb{D}}}
\newcommand{\pr}{{\prime}}
\newcommand{\prr}{{\prime\prime}}
\newcommand{\prrr}{{\prime\prime\prime}}
\title{On Coupled Delayed Van der Pol-Duffing oscillators}
\author{Ankan Pandey\\
SN Bose National Centre for Basic Sciences \\
JD Block, Sector III, Salt Lake \\ Kolkata 700098,  India \\
\and
Mainak Mitra \\ Ramakrishna Mission Residential College (Autonomous) \\ Narendrapur, Kolkata-700103 \\
West Bengal,India \\
\and
A Ghose-Choudhury\footnote{E-mail aghosechoudhury@gmail.com}\\
Department of Physics, Surendranath  College,\\ 24/2 Mahatma
Gandhi Road, Calcutta 700009, India\\
\and
Partha Guha\footnote{E-mail: partha@bose.res.in}\\
SN Bose National Centre for Basic Sciences \\
JD Block, Sector III, Salt Lake \\ Kolkata 700106,  India \\
}

\date{ }

 \maketitle

\smallskip

\smallskip

\begin{abstract}
\textit{We investigate the dynamics of a delay differential coupled Duffing-Van der Pol oscillator equation. Using the Lindstedt's method, we derive the  in-phase mode solutions and then  obtain the slow flow equations governing the stability of the in-phase mode by employing the two variable perturbation method. We solve the slow flow equations using series expansion and obtain conditions for Hopf bifurcation and studied stability of the in-phase mode. Finally, we studied stability and bifurcations of the origin. Our interest in this system is due to the fact that it is related to the coupled laser oscillators.}
\end{abstract}

\smallskip

\paragraph{Mathematics Classification (2010)}:34C14, 34C20.

\smallskip


\section{Introduction}

In this paper we study the dynamics of coupled Duffing-Van der Pol oscillators with delay coupling. The problem of two Van der Pol oscillators with the delay coupling was investigated by Rand and Wirkus in \cite{Wirkus2} who  chose the delay coupling in the damping terms because this form of coupling occurs in radioactively coupled microwave oscillator arrays. The coupling of microwave oscillators via
delayed velocity coupling has been extensively studied by  electrical engineers \cite{Lynch,Lynch2,York,York2}. The two principal features of microwave oscillators are negative resistance and gain saturation.
The former one causes the amplitude of the oscillations to grow while the latter  limits the amplitude of the oscillations. A similar phenomenon also occurs in laser physics and is manifested in the form of relaxation oscillations \cite{Sargent}. As a consequence the Van Der Pol oscillator is often considered as a generic
microwave oscillator.
\smallskip
The present work introduces a cubic nonlinearity effect into the coupled delayed Van Der Pol system considered in \cite{GR}. This leads to a  Duffing-Van Der Pol equation and  provides an important mathematical model for dynamical systems having a single unstable fixed point, along with a stable limit cycle. Examples of such phenomena arise quite naturally in engineering problems. Specifically we analyse two coupled Duffing-Van Der Pol oscillators with delayed velocity coupling. The work is in the spirit of Rand and his collaborators. Time delay is included explicitly in the differential equations rather than the delay in the averaged equations. In the study of two weakly delayed coupled Van der Pol oscillators Wirkus and Rand \cite{Wirkus,Wirkus2} found that both the in-phase and out of phase modes were stable for delays of about a quarter of the uncoupled period of the oscillators. In this paper we generalize the result of R. Rand to two weakly coupled Duffing-Van Der Pol oscillators in which the coupling term involves a time delay $\tau$. We use the method of averaging to obtain the approximate simplified system and then investigate the stability and bifurcation of their equilibria which correspond to periodic motions in the original system. 
 Delay differential equations, it will be recalled, are often used as modelling tools in several areas of applied mathematics, including the study of epidemics, age-structured population growth, automation, traffic flow and problems related to the engineering of high-rise buildings for earthquake protection.
\section{Coupled Van der Pol-Duffing delayed oscillators}
The system of coupled delayed Van der Pol-Duffing oscillators we consider are given by
\be\label{C1a} \ddot{x}_1+\epsilon(x_1^2-1)\dot{x}_1+x_1-\epsilon x_1^3=\alpha\epsilon \dot{x}_2(t-T),\ee
\be\label{C1b} \ddot{x}_2+\epsilon(x_2^2-1)\dot{x}_1+x_2-\epsilon x_2^3=\alpha\epsilon \dot{x}_1(t-T).\ee
The in-phase mode of this coupled system occurs when $x_1=x_2$ for which we have the common equation
\be\label{C2} \ddot{y}+\epsilon(y^2-1)\dot{y}+y-\epsilon y^3=\alpha\epsilon \dot{y}(t-T).\ee
To find a periodic solution of this equation we use the Lindstet perturbation technique and set $\tau=\omega t$ where $\omega=1+\epsilon k+O(\epsilon^2)$. This gives the equation
\be\label{C3}\omega^2y^{\prime\prime}+\epsilon \omega(y^2-1)y^\prime+y-\epsilon y^3=\alpha\epsilon \omega y^\prime(\tau-\omega T),\ee where the primes denote differentiation with respect to the variable $\tau$.
Next we expand $y$ is a power series in $\epsilon$ i.e., set
$$y=y_0+\epsilon y_1+\cdots$$ and equate the coefficients of various powers of $\epsilon$ to get
\be\label{C4} y_0^{\prime\prime}+y_0=0\ee
\be\label{C5}y_1^{\prime\prime}+y_1=-2ky_0^{\prime\prime}-(y_0^2-1)y_0^\prime +y_0^3+\alpha y_0^\prime(\tau-T).\ee
Assuming a solution of the form $y_0=R\cos\tau$ of (\ref{C4}) we substitute this into (\ref{C5}) to obtain
\be\label{C6} y_1^{\prime\prime}+y_1=(2kR+\alpha R \sin T +\frac{3}{4}R^3)\cos\tau-
(R+\alpha R\cos\tau-\frac{1}{4}R^3)\sin\tau+\frac{R^3}{4}\cos 3\tau +\frac{R^3}{4}\sin 3\tau\ee
By demanding the secular terms to vanish we obtain the solutions for the amplitude $R$ and $k$ as given below
\be \label{C7} R=2\sqrt{1+\alpha \cos T},\;\;\;k=-\frac{1}{2}(\alpha \sin T+3(1+\alpha\cos T).\ee
 Consequently for the in-phase mode we have
 \be\label{C8}y\approx y_0=2\sqrt{1+\alpha \cos T}\cos\{1-\frac{\epsilon}{2}(\alpha \sin T+3(1+\alpha\cos T))t\}.\ee

\subsection{Stability of the in-phase mode}

In order to study the stability of the in-phase mode we set
$$x_1=y(t)+w_1, \;\;\;x_2=y(t)+w_2,$$
and linearise the system (\ref{C1a})-(\ref{C1b}) about $w_1=w_2=0$. This leads to the following system of coupled linear delayed differential equations (DDE), namely
\be\label{DD1a}\ddot{w}_1+\epsilon(y^2-1)\dot{w}_1+(1+\epsilon(2y\dot{y}-3y))w_1=\alpha \epsilon\dot{w}_2(t-T),\ee
\be\label{DD1b}\ddot{w}_2+\epsilon(y^2-1)\dot{w}_2+(1+\epsilon(2y\dot{y}-3y))w_2=\alpha \epsilon\dot{w}_1(t-T).\ee
The above system is easily decoupled by the transformation
$$z_1=w_1+w_2, \;\;\;z_2=w_1-w_2,$$
whence they become
\be\label{DD2a}\ddot{z}_1+\epsilon(y^2-1)\dot{z}_1+(1+\epsilon(2y\dot{y}-3y))z_1=\alpha \epsilon\dot{z}_1(t-T),\ee
\be\label{DD2b}\ddot{z}_2+\epsilon(y^2-1)\dot{z}_2+(1+\epsilon(2y\dot{y}-3y))z_2=-\alpha \epsilon\dot{z}_2(t-T).\ee
The decoupled system given in the last two equations have the generic form
\be\label{D3}\ddot{u}+\epsilon(y^2-1)\dot{u}+(1+\epsilon(2y\dot{y}-3y))u=\beta\alpha \epsilon\dot{u}(t-T),\ee
where $u=z_1$ for $\beta=1$ and $u=z_2$ for $\beta=-1$.

\subsection{Two variable perturbation method for Eqn (\ref{D3})}
To investigate the presence of different time scales in the delayed equation (\ref{D3}) we take recourse to the two variable perturbation method involving the times scales $\tau=\omega t$ and $\eta=\epsilon t$. It follows that
$$\frac{du}{dt}=\omega u_\tau+\epsilon u_\eta$$
$$\frac{d^2u}{dt^2}=\omega^2u_{\tau\tau}+2\omega\epsilon u_{\tau\eta}+\epsilon^2u_{\eta\eta}$$
Inserting these expressions into (\ref{D3}) we obtain
\be\label{D4}(\omega^2u_{\tau\tau}+2\omega\epsilon u_{\tau\eta}+\epsilon^2u_{\eta\eta})+\epsilon(y^2-1)(\omega u_\tau+\epsilon u_\eta)+(1+\epsilon(2yy_\tau -3y))u=\epsilon\alpha\beta(\omega u_\tau+\epsilon u_\eta)(\tau-\omega T, \eta-\epsilon T).\ee
Expanding $u$ and $\omega$ in a power series in $\epsilon$, \textit{viz}
$$u=u_0+\epsilon u_1+O(\epsilon^2), \;\;\;\omega=1+\epsilon k+ O(\epsilon^2)$$
 we have upon inserting the expression for $k$ using (\ref{C7}) and (\ref{C8}) and  retaining terms up to the first-order in $\epsilon$:
 \be\label{D5}u_{0\tau\tau}+u_0=0\ee
 $$u_{1\tau\tau}+u_{1}=-2u_{0\tau\eta}+(\alpha\sin T+3(1+\alpha\cos T))u_{0\tau\tau}+(1-4(1+\alpha\cos T)\cos^2\tau)u_{0\tau}$$
\be\label{D6}+[8(1+\alpha\cos T)\cos\tau\sin\tau+6\sqrt{1+\alpha\cos T}\cos\tau]u_0
+\alpha\beta u_{0\tau}(\tau-T, \eta-\epsilon T)\ee
Next assume that (\ref{D5}) has the solution
$$u_0(\tau, \eta)=A(\eta)\cos\tau+B(\eta)\sin\tau$$ so that the delayed term has the form
$$u_0(\tau-T, \eta-\epsilon T)=A_d\cos(\tau-T)+B_d\sin(\tau-T)$$
where $A_d=A(\eta-\epsilon T)$ and $B_d=B(\eta-\epsilon T)$ respectively.
 Inserting  the above solution into (\ref{D6}) and requiring that the secular terms vanish gives the following system of equations for the amplitudes \textit{viz}
\be\label{D7a} \frac{dA}{d\eta}=-\bigg(1+\frac{3\alpha\cos T}{2}\bigg)A+\frac{\alpha\sin T}{2}B+\frac{\alpha\beta\cos T}{2}A_d-\frac{\alpha\beta\sin T}{2}B_d,\ee
\be\label{D7b}\frac{dB}{d\eta}=\bigg(3(1+\alpha\cos T)-\frac{\alpha\sin T}{2}\bigg)A-\frac{\alpha\cos T}{2}B
+\frac{\alpha\beta\sin T}{2}A_d+\frac{\alpha\beta\cos T}{2}B_d.\ee
The last two equations represent the \textit{slow flow} system of delay differential equations.
 The first step to analysing the system (\ref{D7a})-(\ref{D7b}) consists in setting $A=Pe^{\lambda\eta}, B=Qe^{\lambda\eta}$ and $A_d=Pe^{\lambda(\eta-\epsilon T)}, B_d=Qe^{\lambda(\eta-\epsilon T)}$ in (\ref{D7a}) and (\ref{D7b}) which gives
\be \lambda A=-A\bigg(1+\frac{3\alpha\cos T}{2}\bigg)A+\frac{\alpha\sin T}{2}B+\frac{\alpha\beta\cos T}{2}Ae^{-\lambda\epsilon T}-\frac{\alpha\beta\sin T}{2}B e^{-\lambda\epsilon T} \nn ,\ee
\be \lambda B=\bigg(3(1+\alpha\cos T)-\frac{\alpha\sin T}{2}\bigg)A-\frac{\alpha\cos T}{2}B+\frac{\alpha\beta\sin T}{2}e^{-\lambda \epsilon T}A+\frac{\alpha\beta\cos T}{2}e^{-\lambda\epsilon T}B. \nn \ee
Considering the case $\beta=-1$ we are led therefore to the homogeneous system of equations:
$$\left(\begin{array}{ccc}
-\lambda-1-\frac{\alpha\cos t}{2}(3+e^{-\lambda\epsilon T}) & \frac{\alpha\sin T}{2}(1+e^{-\lambda\epsilon T})\\
3(1+\alpha\cos T)-\frac{\alpha\sin T}{2}(1+e^{-\lambda\epsilon T}) & -\lambda-\frac{\alpha\cos T}{2}(1+e^{-\lambda\epsilon T})\end{array}\right)
\left(\begin{array}{ccc} A\\ B\end{array}\right)=\left(\begin{array}{ccc} 0 \\ 0\end{array}\right)$$
For non-trivial solutions it is necessary that the determinant vanishes. We set $\lambda=i\Omega$ and equate the determinant to zero to obtain
$$-\Omega^2+i\Omega \bigg(1+\alpha\cos T(2+e^{-i\epsilon\Omega T})\bigg)+\frac{\alpha^2\cos^2 T}{4}
(3+4e^{-i\epsilon\Omega T}+e^{-2i\epsilon\Omega T})$$
$$+\frac{\alpha^2\sin^2 T}{4}(1+2e^{-i\epsilon\Omega T}+e^{-2i\epsilon\Omega T})+
\frac{\alpha\cos T}{2}(1+e^{-i\epsilon\Omega T})-\frac{3}{2}\alpha\sin T(1+\alpha\cos T)e^{-i\epsilon\Omega T}=0$$
 We separate the real and imaginary parts of the last equation to get
%Imaginary part:
%$$\Omega+\frac{\alpha\cos T}{2}\left(\Omega(1+\cos\epsilon\Omega T)-\sin\epsilon\Omega T\right)+\Omega\frac{\alpha\cos %T}{2}(3+\cos\epsilon\Omega T)-\frac{\alpha^2\cos^2T}{4}(4\sin\epsilon\Omega T+\sin2\epsilon\Omega T)$$
%$$-\frac{\alpha^2\sin^2 T}{4}(2\sin\epsilon\Omega T+\sin2\epsilon\Omega T)-\frac{3\alpha\sin T}{2}(1+\alpha\cos %T)\sin\epsilon\Omega T=0$$
% and for the Real part:
% $$-\Omega^2+\frac{\alpha\cos T}{2}(1+\cos\epsilon\Omega T+\Omega \sin\epsilon\Omega T)+\frac{\alpha\cos %T}{2}\Omega\sin\epsilon\Omega T+\frac{\alpha^2\cos^2 T}{4}(3+4\cos\epsilon\Omega T+\cos2\epsilon\Omega T)$$
% $$+\frac{\alpha^2\sin^2 T}{4}(1+2\cos\epsilon\Omega T+\cos2\epsilon\Omega T)+\frac{9}{4}(1+\alpha^2\cos^2 T %+2\alpha\cos T)+\frac{3\alpha\sin T}{2}(1+\alpha\cos T)(1+\cos\epsilon\Omega T)=0$$
%We expand $T$ and $\Omega$ in a power series in $\epsilon$, viz,
%$$T=T_0+\epsilon T_1+\epsilon^2 T_2+\cdots$$
%$$\Omega=\Omega_0+\epsilon\Omega_1+\epsilon^2\Omega_2+\cdots$$
%Substituting these expansions into the above expression and  separating the imaginary and real parts respectively,\\
%\underline{Coefficient of $\epsilon^0$:}\\
%$$\Omega_0(1+3\alpha\cos T_0)=0\Longrightarrow \;\;\cos T_0=-\frac{1}{3\alpha}$$
%$$\Omega_0=\frac{1}{3}\sqrt{3p^2-3p-1}, \;\;\;p=\sqrt{9\alpha^2-1}$$
%\underline{Coefficient of $\epsilon^1$:}\\
%$$ T_1=-\frac{T_0}{9}(p-3)$$
%$$\Omega_1=-\frac{T_0}{108}\left[\frac{3p^3-5p^2-6p-24}{\sqrt{p^2-3p-1}}\right],$$

\underline{Real Part:}
$$-\Omega^2+\frac{\Omega\alpha\cos T}{2}\sin\Omega\epsilon T+\frac{\alpha\cos T}{2}+\frac{3\alpha^2}{4}\cos^2T+\frac{\alpha^2\cos^2T}{4}\cos\Omega\epsilon T$$ $$+\frac{\alpha\cos T}{2}\cos\Omega\epsilon T+\frac{\alpha\cos T}{2}\Omega\sin\Omega\epsilon T+\frac{3\alpha^2\cos^2T}{4}\cos\Omega\epsilon T+\frac{\alpha^2\cos^2T}{4}\cos 2\Omega\epsilon T$$
\be\label{w3}+\frac{\alpha^2\sin^2T}{4}(1+\cos 2\Omega\epsilon T+2\cos\Omega\epsilon T)-3(1+\alpha\cos T)\frac{\alpha\sin T}{2}(1+\cos\Omega\epsilon T)=0.\ee\\
\underline{Imaginary Part:}
$$\Omega+\frac{\Omega 3\alpha}{2}\cos T+\frac{\Omega\alpha\cos T}{2}\cos\Omega\epsilon T+\frac{\alpha\cos T}{2}\Omega-\frac{\alpha^2\cos^2T}{4}\sin\Omega\epsilon T$$ $$+\frac{\alpha\Omega\cos T}{2}\cos\Omega\epsilon T-\frac{\alpha\cos T}{2}\sin\Omega\epsilon T-\frac{3\alpha^2\cos^2T}{4}\sin\Omega\epsilon T-\frac{\alpha^2\cos^2T}{4}\sin 2\Omega\epsilon T$$
\be\label{w4}-\frac{\alpha^2\sin^2T}{4}\sin 2\Omega\epsilon T+\frac{\alpha^2\sin^2T}{4}(-2\sin\Omega\epsilon T)+3(1+\alpha\cos T)\frac{\alpha\sin T}{2}\sin\Omega\epsilon T=0.\ee\\

We expand $T$ and $\Omega$ in a power series in $\epsilon$, viz,
$$T=T_0+\epsilon T_1+\epsilon^2 T_2+\cdots$$
$$\Omega=\Omega_0+\epsilon\Omega_1+\epsilon^2\Omega_2+\cdots$$

Substituting these expansions into the equations (\ref{w3}) and (\ref{w4}) and equating terms of the various orders of $\epsilon$ we obtain,\\

\underline{Coefficient of $\epsilon^0$:}\\
\be\label{w3}\Omega_0(1+3\alpha\cos T_0)=0\Longrightarrow \;\;\cos T_0=-\frac{1}{3\alpha}\ee
\be\label{w3}\Omega_0=\frac{1}{3}\sqrt{p^2-6p-1}, \;\;\;p=\sqrt{9\alpha^2-1}\ee
\underline{Coefficient of $\epsilon^1$:}\\
\be\label{w4}T_1=\frac{T_0}{3}\left(1-\frac{p}{3}\right) \;\;\;\;\;\;or   \;\;\;\;\;\; T_1=-\frac{T_0}{9}\left(p-3\right)\ee\\
\be\label{w6}\Omega_1=-\frac{T_0}{54}\left[\frac{3p^3-7p^2-9p-21}{\sqrt{p^2-6p-1}}\right].\ee
The higher order terms can also be found in a similar manner. The expressions for $\Omega$ and $T$ gives the conditions for the Hopf bifurcation of the solution $y(t)$. If consider only the first order then the condition is given as $\alpha=-\frac{1}{cosT_0}$. To check the stability of the limit cycle, originated out of this Hopf bifurcation, let $\lambda=R+i\,\Omega$ and expand $\lambda$ in the neighbourhood of the limit cycle. As the real part $R$ would be zero at the cycle, The stability could be checked by noticing the sign of the first derivative. From the calculations, it was found that the first order term is given as  $R=\frac{1}{2}(-1-3\,\alpha\,cos T)$. This expression is expected as it reduces to the Hopf condition for $R=0$. However, one point to keep in mind is that these conditions are not exact and true only for first order and with higher orders, corrections will be added to the expressions and more precise conditions could be obtain as shown by Gluzman and Rand in \cite{GR}.\\
 The saddle-node condition could be obtain by assuming $\lambda=0$ and the determinant vanishes. For $\beta=-1$ the conditions are
\be \alpha=0,\,\,\&\,\,\alpha=-\frac{2(cos T - 3\,sin T)}{3 + cos 2T - sin 2T}.\nn \ee
The $\beta=1$ case gives the conditions for the birth of in-phase periodic mode , corresponding to the independence of $A$ and $B$ on $\eta$. The condition is given as
\be \alpha=\frac{-1}{cos T}. \nn \ee
 There are no correction terms involved in this expression as we considered the case for $\lambda=0$. Also, there is no $\epsilon$ dependence, which is evident as these are slow flow equations and only dependence on $\epsilon$ is in the delay. In the next section, we will see that this condition corresponds to the change in stability of the origin itself.

\section{Stability analysis of velocity delay coupling}
The Duffing-Van der Pol oscillator defined in (\ref{C1a}, \ref{C1b}) possesses an unstable fixed point at origin and a stable limit cycle  around it in uncoupled case. Coupling the system without delay does not affect the limit cycle and it continue to exist for all parameter values. However, with delay coupling, the stability of the origin changes for some parameter arrangements. To study the stability of origin, assume the variation of the coordinates in the neighbourhood of the origin proportional to $e^{m\,t}$ which gives the characteristic equation for the stability of origin as

\be m^2 + \epsilon m + 1 = \alpha \epsilon\, m \,e^{-m T}. \label{eq:genchar} \ee

The characteristic equation is transcendental in nature and could not have closed form solutions. To analyse it for periodic orbit, consider the $\lambda=i\,m_I$ which on solving gives the delay curves as
\be m_{I\pm}=\bigg[ \frac{1}{2}(\epsilon^2(\alpha^2 - 1) + 2) \pm \frac{1}{2}((\epsilon^2(\alpha^2 - 1) + 2)^2 - 4)^{1/2}\bigg]^{1/2}. \nn \ee
and satisfies
\be cos(m_\pm T_\pm)=\frac{-1 }{\alpha}, \label{eq:Veloorigindelay}\ee

\begin{figure}
\centering
\includegraphics[bb = 0 0 500 500, width=8cm, height=8cm, keepaspectratio]{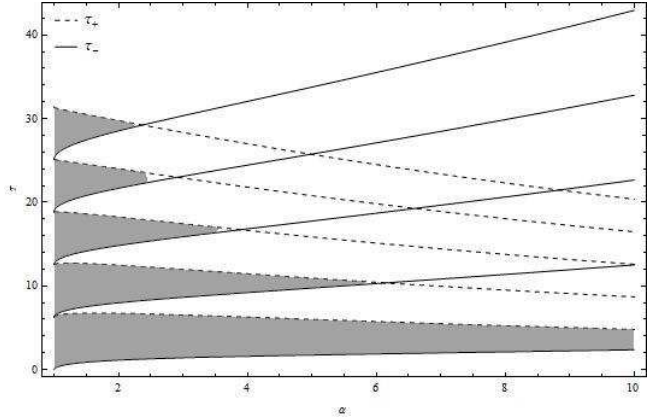}
\caption{The plot shows the variation of delay with coupling coefficient for values $\omega=1.0,\,k=0.1$, with $n=\{1,2,3,4\}$ for $T_+$ and $n=\{1,2,3\}$ for $T_-$.}
\label{fig:OriginVdAd}
\end{figure}

Condition in (\ref{eq:Veloorigindelay}) gives the condition for Hopf Bifurcation of the origin.
The delay curves could be expanded for small $\epsilon$ and approximated as
\be  m_{I\pm}=1+\frac{1}{2}\sqrt{\alpha^2 - 1}\, \epsilon + O(\epsilon^2), \ee

which, for the first order, gives $\alpha=-\frac{1}{cos T}$. This is what we got from the slow flow analysis of the periodic solution. The higher order corrections were not present in the slow flow, however the stability analysis of the origin provides some higher order corrections. \\
Now, the characteristic equation (\ref{eq:genchar}), in a more general form, had been studied and analysed before and given the conditions 
\be \alpha < 1 \,\,\,\,\&\,\,\,\,\alpha^2 > 1, \nn  \label{eq:VdOriginSC} \ee
the conclusions of the studies are given in the following theorem
\begin{theorem}
For $T_\pm(n)$ defined as (\ref{eq:Veloorigindelay}), there is a positive integer $n$ such that there are $n$ switches from stability to instability to stability, i.e., when
\be T \in [0,T_+(0)]\cup(T_-(0),T_+(1))\cup...\cup(T_-(n-1),T_+(n)) \nn \ee
all roots of (\ref{eq:genchar}) have negative real parts.
\end{theorem}
 The theorem is part of \cite{CG} and the proof is straightforward, for which please refer to \cite{BH}. The theorem provides the conditions for the real part of the roots of (\ref{eq:genchar}) to be negative which denotes the stability of the origin. The origin is a stable focus for $\alpha<-1$ for some sets of delay values as specified in theorem (3.1). For $-1<\alpha<1$ the eigenvalues are purely real and the upper bound of its values is obtain as the roots of the characteristic equation for $T=0$, given as
\be m_\pm=-\frac{1}{2}\epsilon(1-\alpha)\pm \frac{1}{2}\sqrt{\epsilon^2(1-\alpha)^2-4}. \ee
The upper boundedness could be proven by Rouche's theorem. Clearly, the origin is unstable in this region.\\
The origin for $\alpha>1$ shows transition as in the case for $\alpha<-1$. It is stable for some sets of delay values and unstable for other. The unstable phases corresponds with the stable limit cycle. Figure (\ref{fig:OriginVdAd}) shows the stability regions (shaded) for $\alpha>1$ case.

\section{Numerical Simulation}

\begin{figure}
\centering
\subfloat[$T=2.7,\,\,\alpha=1.5$]{
\includegraphics[bb = 0 0 500 500, width=8cm,height=8cm,keepaspectratio]{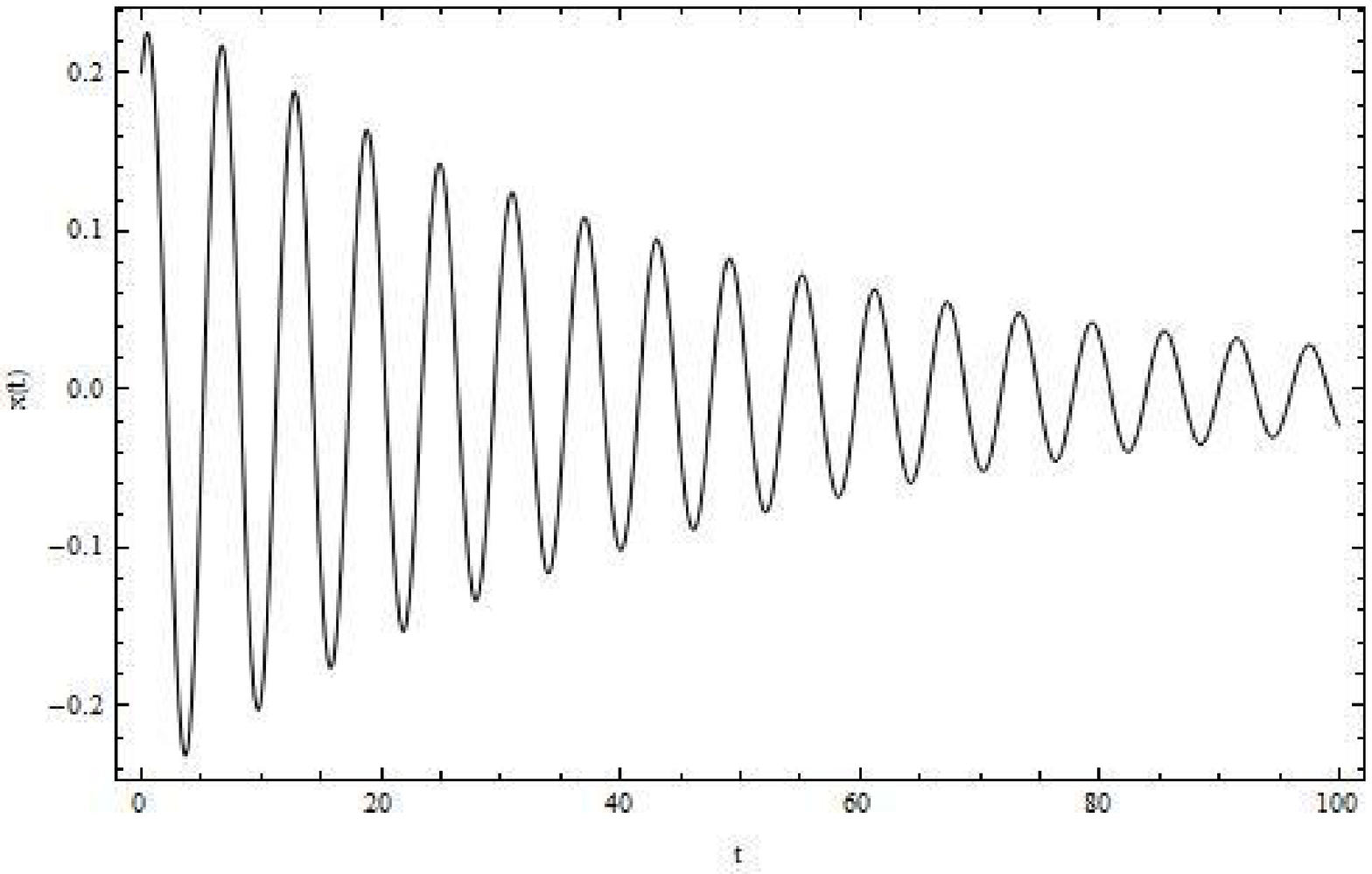}}
\subfloat[$T=7.2,\,\,\alpha=1.5$]{
\includegraphics[bb = 0 0 500 500, width=8cm,height=8cm,keepaspectratio]{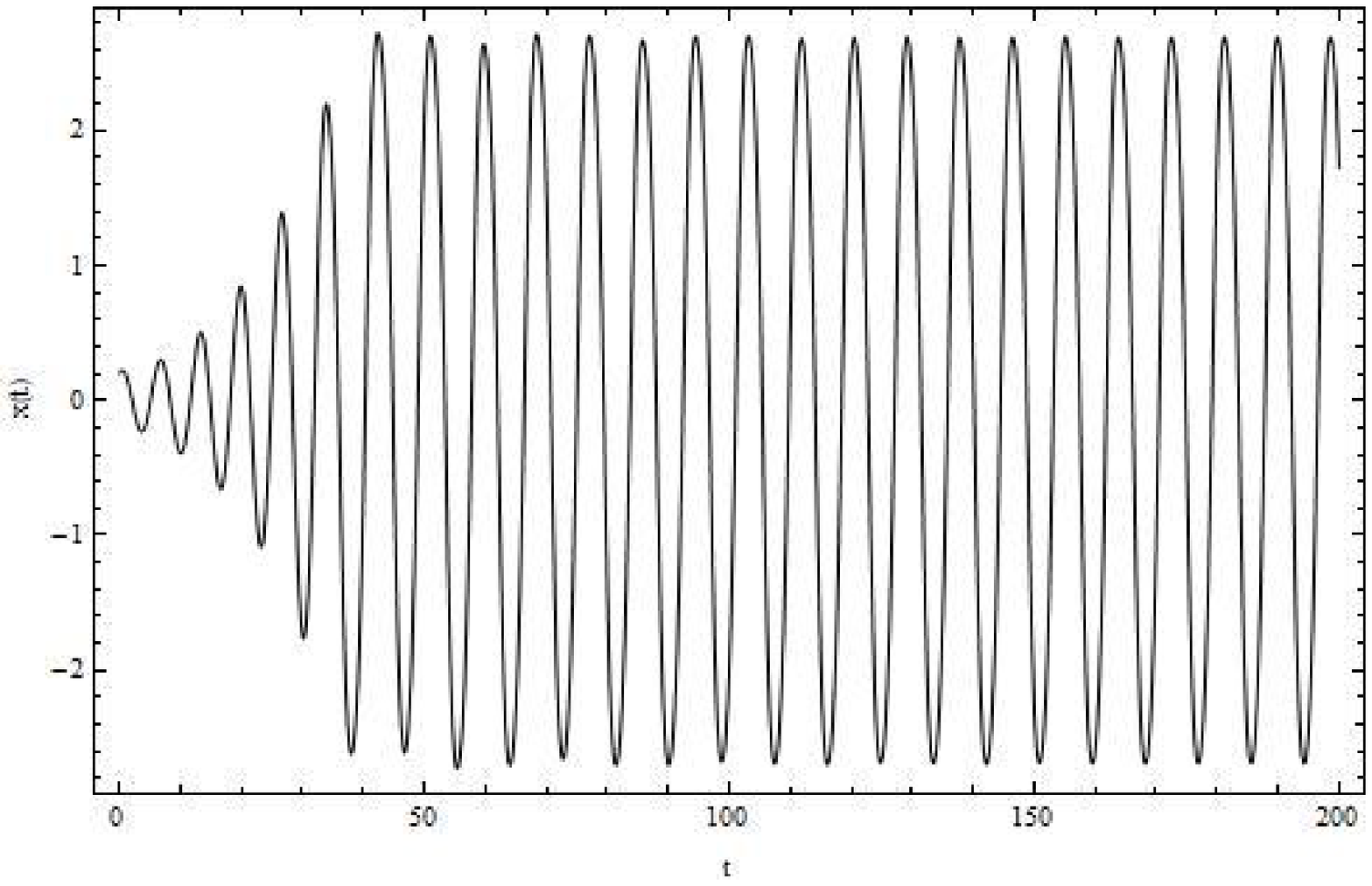}}
\caption{The plot shows time series of system (\ref{C1a},\ref{C1b}) for parameters $\omega=1.0,\,k=0.1,\,\lambda=-0.1,\,\,\alpha=1.5$ and $T=2.7,\,\,\&\,\,T=7.2$ denoting the stability of origin and limit cycle, respectively.}
\label{fig:OriginVdAdTimeseries}
\end{figure}

In previous sections, we derived the conditions for the Hopf bifurcation of the origin which corresponds with the birth of in-phase periodic solution. Further, we have seen that the origin is stable for $\alpha^2>1$ which corresponds to the case of amplitude death of the oscillator. In this section, we numerically simulate the coupled system.\\
From theorem 4.1 we have conditions for the stability of the origin. We simulated the system for $\alpha=1.5,\,\,\&\,\,\epsilon=0.1$ with initial conditions $x(0)=0.5,\,\,y(0) = 0.5,\,\,\dot{x}(0) = 0.5,\,\,\dot{y}(0) = 0.5$. The corresponding critical delay values are $T \in [0,0.7953]\cup[0.8894,6.7371]\cup[7.5336,12.6789]$. The resultant plots are shown in figure (\ref{fig:OriginVdAdTimeseries}), which clearly shows the expected behaviour and in line with analytical calculations.

\section{Conclusion}

In this paper we studied coupled Duffing-Van der Pol oscillators by velocity delay terms. At first 
we used Lindstedt-Poincar\'e method to obtain an approximate expression for the in-phase mode. Then we
studied the stability of the in-phase mode by applying the two variable perturbation method to
$$\ddot{u}+\epsilon(y^2-1)\dot{u}+(1+\epsilon(2y\dot{y}-3y))u=\beta\alpha \epsilon\dot{u}(t-T),$$
where $u=z_1$ for $\beta=1$ and $u=z_2$ for $\beta=-1$.
This resulted in the DDE slow flow. This resulted in a system of modified ODEs which possessed Hopf and saddle-node bifurcations. The vanishing of 
the determinant of the slow flow of DDE yields the  nontrivial solution, but it is harder to solve. So, motivated from Gluzman-Rand work, we sought a series solution. Further, the stability of the in-phase mode was analysed. This stability corresponds to change in stability of the origin which was shown in section 4. The coupled system was also numerically studied and it was observed that numerical results are in sync with the analytical calculations.

%\newpage
%\section*{Figure captions}
%Figure (\ref{fig:OriginVdAd}) : The plot shows the variation of %delay with coupling coefficient for values $\omega=1.0,\,k=0.1$, with $n=\{1,2,3,4\}$ for $T_+$ and $n=\{1,2,3\}$ for $T_-$. \\
%Figure (\ref{fig:OriginVdAdTimeseries}) : The plot shows time series of system (\ref{C1a},\ref{C1b}) for parameters $\omega=1.0,\,k=0.1,\,\lambda=-0.1,\,\,\alpha=1.5$ and $T=2.7,\,\,\&\,\,T=7.2$ denoting the stability of origin and limit cycle, respectively.

\end{document}